\documentclass[
prl,
reprint,
amsmath,
amssymb,
]{revtex4-1}
\usepackage{graphicx}
\begin{document}

\title{Density dependent exciton dynamics and L valley anisotropy in germanium}

\author{M.~\surname{Stein}}
\email{markus.stein@physik.uni-marburg.de}
\author{C.~\surname{Lammers}}
\author{P.~\surname{Springer}}
\altaffiliation{Current address: Institute of Technical Physics, German Aerospace Center, Paffenwaldring 38-40, 70569 Stuttgart, Germany}
\author{P.-H.~\surname{Richter}}
\author{S.\,W.~\surname{Koch}}
\author{M.~\surname{Kira}}
\author{M.~\surname{Koch}}
\affiliation{Department of Physics and Material Sciences Center, Philipps-Universit\"{a}t Marburg, Renthof 5, 35032 Marburg, Germany}

\date{\today}

\begin{abstract}
Optical pump-THz probe spectroscopy is used to investigate the exciton formation dynamics and its intensity dependence in bulk Ge. Associated with the intra-excitonic 1s-2p transition, the gradual build-up of an absorption peak around 3.1\,meV (0.75\,THz) signifies the delayed exciton formation after optical pump which is accelerated for   
higher excitation densities. Analyzing the spectral shape of this THz absorption resonance, two distinct resonances are found which are attributed to the mass-anisotropy of L valley electrons via a microscopic theory.  
\end{abstract}

\pacs{Valid PACS appear here}
                            
                          
\maketitle
  
Electrons and holes in photoexcited semiconductors can form bound pairs of hydrogen-like excitons due to the attractive Coulomb interaction. 
Since decades, techniques like linear absorption \cite{Haug:1619067,Kazimierczuk:2014bn} time-resolved photoluminescence \cite{Feldmann:1987jk,Deveaud:1991gd,Imhof:2011kd}, pump-probe spectroscopy \cite{Knox:1986ba,Heberle:1995fp}, and four-wave-mixing \cite{Wang:1990kd,Koch:1993gu,Cundiff:2009ig} have been used to study these quasi-particles intensively.
As a consequence, most aspects and many important details of the light-matter interaction phenomena in semiconductors are now well understood \cite{Kira:1999bk,Kira:2011fb,Kira:2009bs}.
  
However, despite all these studies, the conditions and dynamical characteristics of exciton formation are still discussed controversially. 
The microscopic analysis of Kira et al. \cite{Kira:1998gd} showed that the formation time of excitons cannot simply be deduced from the build-up of a luminescence peak at the spectral position of the excitonic transition.
As an alternative method, it was suggested to monitor the appearance of the 1s-2p intra-excitonic resonance which 
can only occur if an optically induced coherent excitonic polarization and/or incoherent excitonic populations are present in the material \cite{PhysRevLett.87.176401}.
For most of the popular semiconductor materials, the observation of this resonance requires the use of optical pump and terahertz (THz) probe spectroscopy.
Provided with a suitable time resolution, this technique allows for the investigation of the formation and decay of the characteristic intra-excitonic transitions \cite{koch2006semiconductor}.

In fact, the optical pump and THz probe approach was pioneered already in 1994 by Groeneveld and Grischkowsky \cite{Groeneveld:1994he}, well before the above mentioned theoretical work \cite{Kira:1998gd}. 
Groeneveld and Grischkowsky studied the 1s-2p transition of heavy-hole excitons in GaAs/AlGaAs multiple quantum wells. 
Although the signal-to-noise ratio and the analysis of the results was rather limited at that time, these investigations were the first experiments of this kind. 
About ten years later, similar experiments with much improved resolution were performed by Kaindl et al. \cite{Kaindl:2003iy} and Huber et al. \cite{Huber:2005iu} who observed the build-up and decay of excitonic populations in GaAs/AlGaAs multiple quantum wells. 

In subsequent years, also indirect semiconductors have been studied. 
Suzuki and Shimano investigated the formation of excitons and electron-hole droplets \cite{Suzuki:2009bq} and the cooling dynamics of the optically excited carriers \cite{Suzuki:2011ii} in Si. 
Furthermore, Sekiguchi and Shimano monitored intra-excitonic transitions in Ge focusing in particular on the Mott transition \cite{Sekiguchi:2015kr}. 

Both, Si and Ge show a strong electron mass-anisotropy in their side valleys. 
For Ge, for instance, it is known since the 1970s, that this anisotropy leads to a splitting of the 1s excitonic resonance in the absorption spectrum \cite{Frova:1975jv}. 
To allow for the modeling of the THz response of such indirect semiconductors with anisotropic effective masses, a microscopic theory was developed recently \cite{Springer:2016hz}. 
The numerical studies predict a characteristic modification of the 1s-2p transition resonance which arises from the asymmetric parabolicity of the L valley in the conduction band. 
In particular, 2p excitons energetically split resulting in two resonances in the THz absorption spectrum. 
However, such a double resonance structure has not yet been observed experimentally.

In this paper, we investigate the dynamics of exciton formation in bulk Ge for different excitation densities using optical pump and THz probe spectroscopy. 
We show that the exciton formation occurs faster when the excitation density is increased. 
Additionally, we unambiguously identify the unique double resonance attributed to the L valley mass anisotropy.   
\begin{figure}[!ht]
  \includegraphics[width=8.5cm]{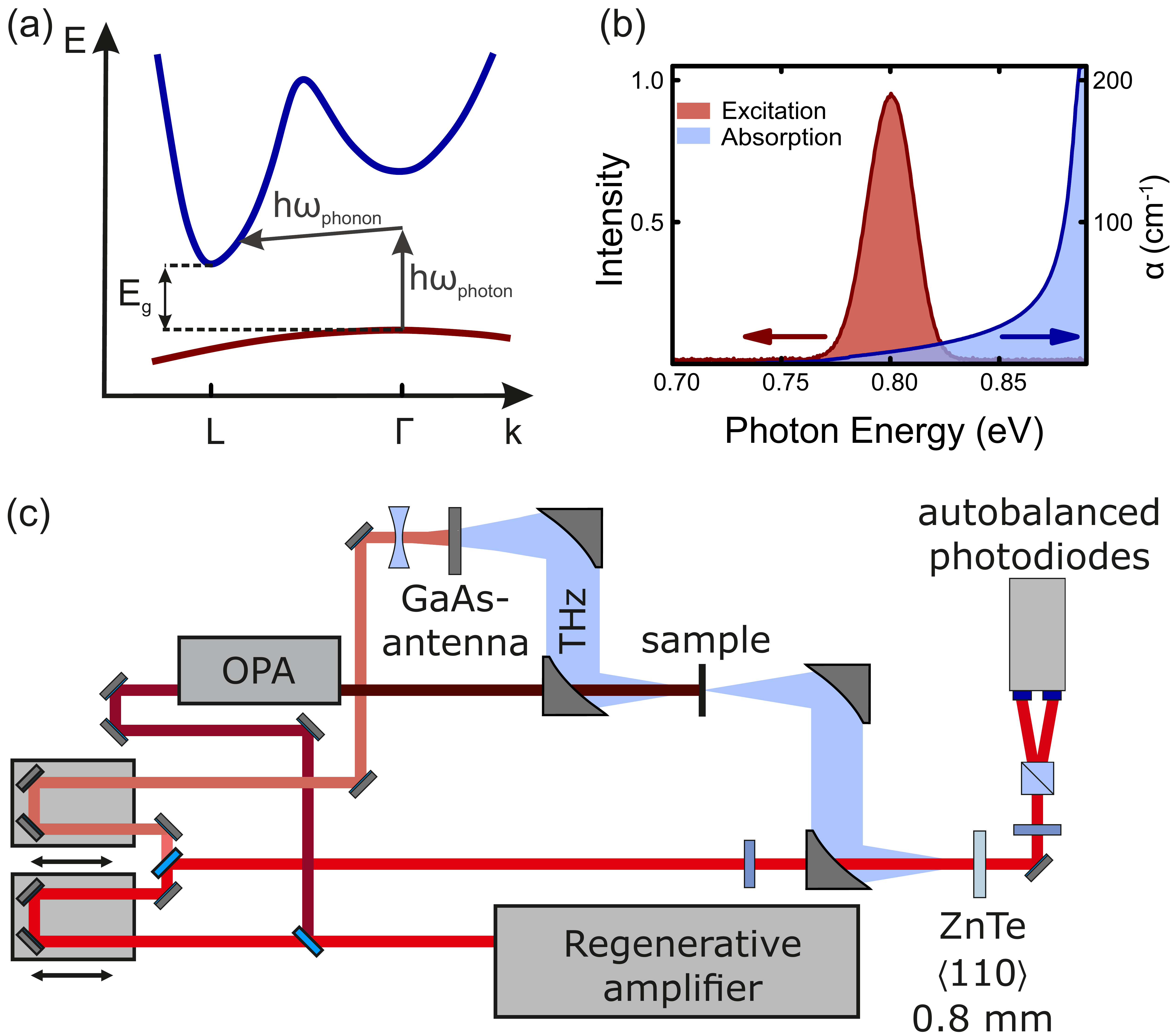}
  \caption{
  (a) The Ge sample is optically excited around 0.8\,eV (red curve). Here, the absorption of Ge (blue curve) is relatively low. (b) A sketch of the indirect band structure of Ge shows an electron mass anisotropy in the L valley. The excitation energy is barely above the indirect band gap. (c) A schematic of the optical pump and THz probe setup.    
  }
  \label{fig:fig1}
\end{figure}

We use a sample of undoped n-type Ge with a thickness of $l$=500\,$\mu$m and a room temperature resistivity larger than 30\,$\Omega$cm\textsuperscript{-1} for our experiments. 
A schematic of the indirect band structure of Ge can be seen in Fig.~\ref{fig:fig1}(a).
The sample is cooled down to 11\,K in a continuous flow liquid-He cryostat. 
A 1\,kHz titanium-sapphire regenerative amplifier system provides 35\,fs-pulses spectrally centered around 800\,nm. 
The pulse train is split into three parts. 
After frequency conversion in an optical parametric amplifier to 0.8\,eV (1550\,nm), the first part optically excites the Ge crystal in the low energy tail of the absorption spectrum (see Fig.~\ref{fig:fig1}(b)). 
As this excitation is only 55\,meV above the indirect band gap the absorption coefficient is low. 
This ensures a high penetration depth of the pump beam and therefore a nearly homogeneous excitation profile in propagation direction.
With a diameter of 3.5\,mm the optical pump spot is laterally larger than the THz probe spot.  
The second part of the pulse train excites a LT-grown large aperture GaAs antenna which emits THz pulses with fields up to 15\,kV/cm \cite{Ewers:2012kz,Drexler:2014cu}.
Field strengths of less than 0.5\,kV/cm are used to ensure that the THz field acts as a probe and does not ionize the excitons \cite{Ewers:2012kz}. 
The third part of the pulse train is used to record the THz probe pulse via electro optical-sampling by means of an 800\,$\mu$m thick ZnTe crystal \cite{Nahata:1996ct}. 
A pair of autobalanced photodiodes are used to sample the waveform of the THz field.
The THz part of our setup is purged with nitrogen in order to get rid of water vapor absorption.
This setup, which is depicted in Fig.~\ref{fig:fig1}(c), can detect THz pulses with a bandwidth ranging from 0.4-12\,meV (0.1-3\,THz) and a sufficiently high signal to noise ratio.
Fourier transformation of the measured waveforms yield the frequency domain fields of the reference pulse $E(\omega)$  as well as its pump induced change $\Delta E(\omega)$. 
The absorption $\alpha (\omega)$ as well as the change of the real part of the dielectric function $\Delta \epsilon (\omega)$ are given by  \cite{Ulbricht:2011fl}
\begin{equation}
	\alpha (\omega)=\frac{2}{l} \operatorname{Re} \left(\frac{\Delta E(\omega)}{E(\omega)} \right),
\end{equation}
\begin{equation}
	\Delta \epsilon (\omega)=\frac{2c_{0} \sqrt{\epsilon_{r}}}{\omega l} \operatorname{Im} \left(\frac{\Delta E(\omega)}{E(\omega)} \right),
\end{equation}
where $c_{0}$ is the speed of light in vacuum and $\epsilon_{r}$ the dielectric constant of the material.
\begin{figure}[!hb]
  \includegraphics[width=8.5cm]{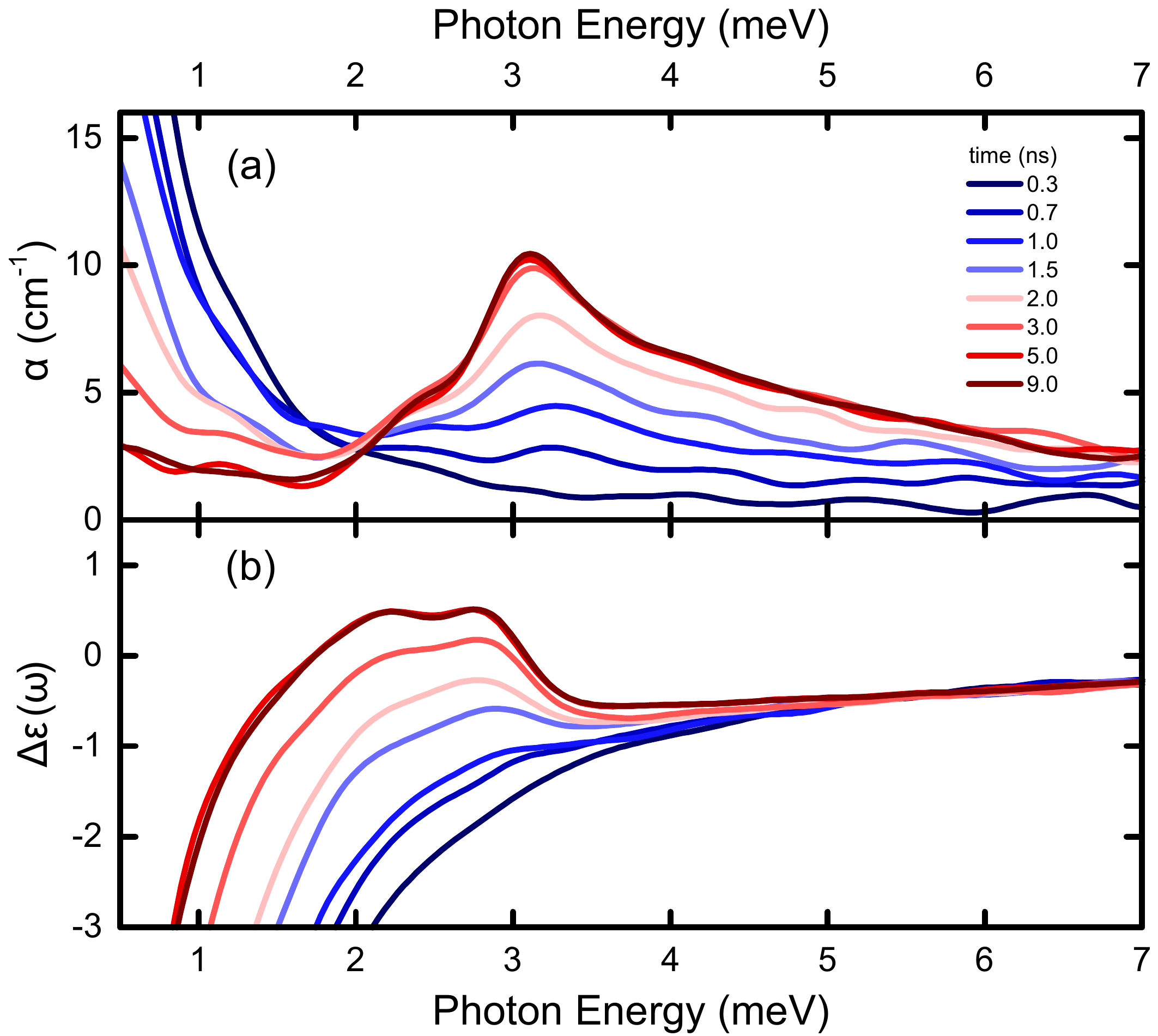}
  \caption{
  (a) THz absorption spectra for different time delays between optical excitation and THz probe pulses. (b) Change of the real part of the dielectric function for the same time delays. 
  }
  \label{fig:fig2}
\end{figure}

To study exciton formation times, it is useful to analyze the THz probe spectra at several instants after the optical pulse has excited the Ge sample, see Fig.~\ref{fig:fig2}(a).
Here, the carrier density is estimated to be  $5 \cdot 10^{14}/\text{cm}^3$, i.e., far below the exciton Mott density in Ge which according to Sekiguchi and Shimano is roughly at $1 \cdot 10^{16}/\text{cm}^3$. We can see in Fig.~\ref{fig:fig2}(a) that shortly after the excitation, the THz absorption spectra are dominated by the response of an electron-hole plasma which gives rise to strong absorption below photon energies of 1.5\,meV. 
The corresponding change of the dielectric function shows a negative response, see Fig.~\ref{fig:fig2}(b), which is characteristic for an electron-hole plasma \cite{Ulbricht:2011fl}.
However, within a few nanoseconds the contribution of the electron-hole plasma vanishes. 
Simultaneously, we observe the formation of excitons as indicated by the occurrence of a prominent absorption peak around 3.1\,meV which is associated with the intra-excitonic 1s-2p transition and agrees with earlier findings \cite{Sekiguchi:2015kr}. 
An additional absorption resonance appears at 2.4\,meV which is accompanied by an oscillatory signature of the dielectric function.

To discuss the exciton dynamics, we notice that
first excitonic signatures are visible as a weak peak in the absorption spectrum around 3.3\,meV for a pump-probe time delay of 0.7\,ns. 
This peak evolves into a clear excitonic resonance reaching 50$\%$ of its final absorption strength around 1.5\,ns after excitation. 
It takes up to 5\,ns till the excitonic peak reaches its final oscillator strength and last features of a significant electron-hole plasma vanish. 
For the temporally adjacent 4\,ns, no evidence of recombination or formation of any further many-particle states like biexcitons or electron-hole droplets can be observed. 

\begin{figure}[!b]
  \includegraphics[width=8.5cm]{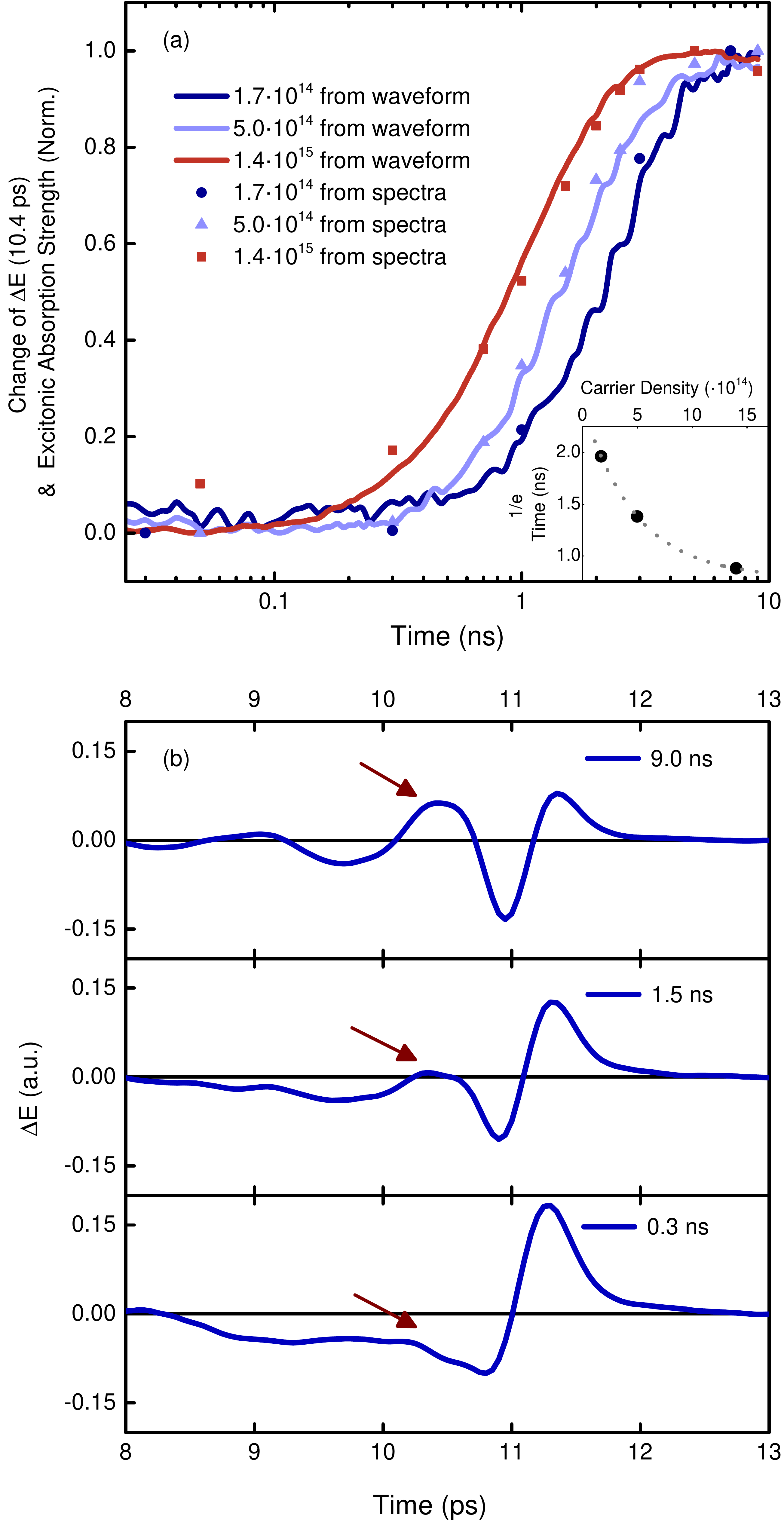}
  \caption{
(a) Rise time of the excitonic response gained by two different methods. Symbols were extracted from the absorption strength of the excitonic resonance, while solid curves are extracted from transients at a temporal positon of 10.4\,ps in the THz pulse. The inset shows build-up times for three different excitation densities obtained by ordinary exponential fits of the transients.
(b) Waveforms of the pump induced change of the probing THz pulses for different times after excitation. The exciton formation can be monitored at certain temporal positions of the waveform. Most noticeable is the change of the waveform at 10.4\,ps, while the system develops from an electron-hole plasma into bound excitons.  
  }
  \label{fig:fig3}
\end{figure}

To quantify this results, we read out the absorption at 3.1\,meV for the different time delays, subtract a weak offset from the plasma response at 50\,ps and finally normalize the data. 
In Fig.~\ref{fig:fig3}(a), we show the temporal evolution of the excitonic absorption strength for different excitation densities (symbols).
Even though it is possible to extract an exciton formation time with this method, the time resolution is limited by the low number of spectra that can be measured within a reasonable time.
To follow the dynamics of the exciton formation with a much better time resolution, we therefore use an alternative experimental route. 
This method exploits a prominent feature in THz waveforms which can be attributed to the build-up of exciton populations.
For illustration purposes, Fig.~\ref{fig:fig3}(b) shows the pump induced change of three THz pulses transmitted through the sample after 0.3, 1.5 and 9\,ns, respectively. 
At a temporal position of 10.4\,ps, $\Delta E$ continuously changes from negative (0.3\,ns delay) to positive (9\,ns delay).
This change is directly correlated to the presence of an exciton population: at early times, when the change is negative, a plasma response dominates the THz absorption spectrum.
With progressing time, when the fraction of carriers bound into incoherent excitons rises, the change increasingly becomes more positive.
Hence, the signal amplitude at this time delay is a fairly good measure for the presence of excitons.
Using this method, Fig.~\ref{fig:fig3}(a) illustrates the normalized change of $\Delta E$ shown as solid lines along with the previous results obtained from the spectral absorption represented by symbols which are in good agreement with one another.
Fitting ordinary exponential functions, we find the characteristic exciton formation times 
ranging from 1.96\,ns for the lowest density to 0.88\,ns for the highest density investigated. 
Hence, we can experimentally verify our previous theoretical predictions in Ref.~\onlinecite{Hoyer:2003cv} that assert an accelerated exciton formation for an increased excitation density. 

Note, that in contrast to the experiments on direct III-V semiconductors \cite{Kaindl:2009ek}, we do not observe any noticeable exciton fraction for early times - neither in the spectra nor in the transients.  
We find here, that the formation of excitons in Ge starts only after ~200-700\,ps depending on the  carrier density. 
This is consistent with the typical cooling dynamics of excited charge carrier systems mediated by acoustic phonons \cite{Kash:1984ej}.

\begin{figure}[!t]
  \includegraphics[width=8.5cm]{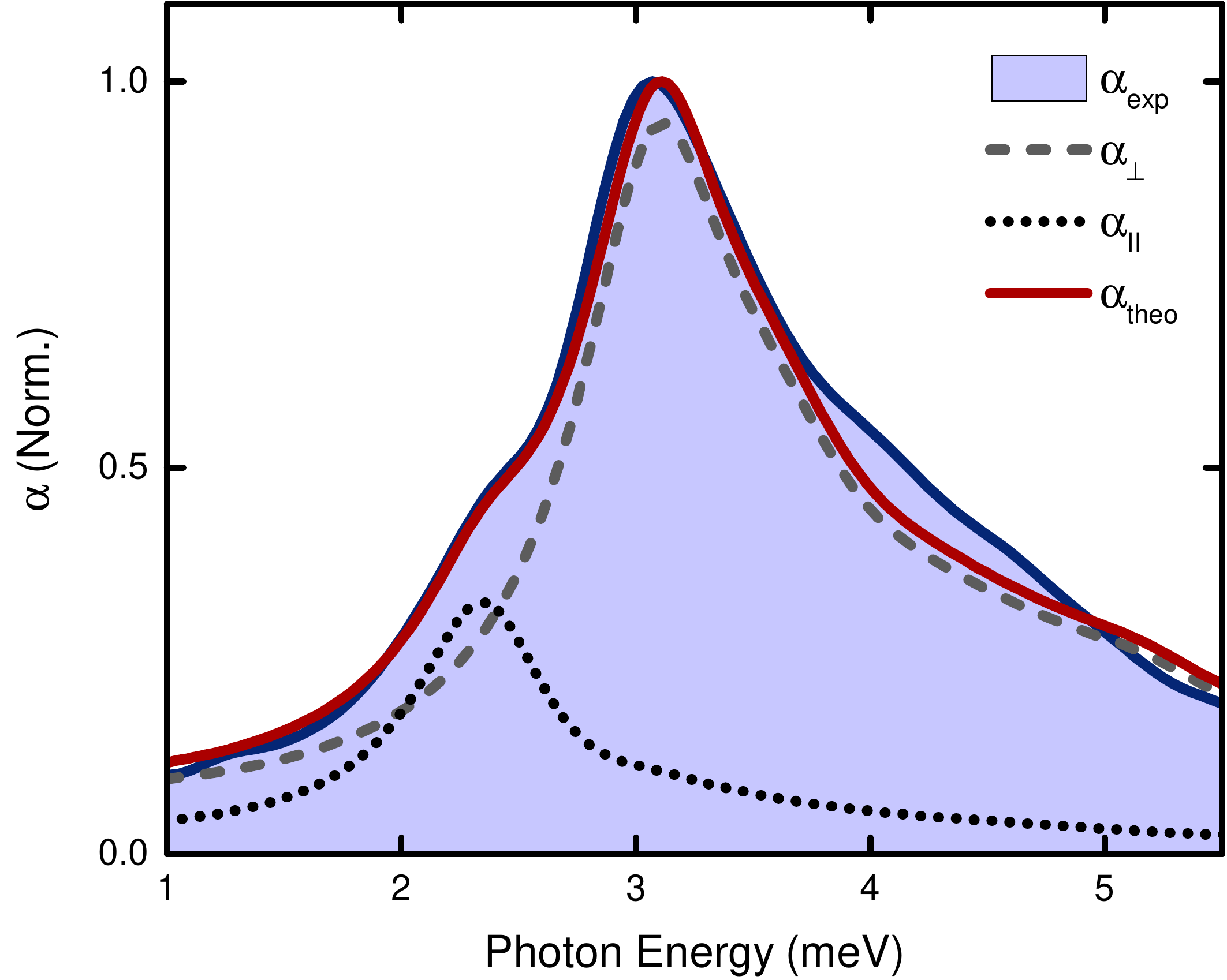}
  \caption{
  Comparison  between the experimental and theoretical results. The measured spectrum (shaded area) was taken 7\,ns after optical excitation. The theoretical spectrum (red) is a superposition of absorption spectra for parallel (dotted) and perpendicular (dashed) polarization.
  }
  \label{fig:fig4}
\end{figure}

Besides experimental extraction of relevant time scales, a comparison to theoretically predicted absorption line shapes was performed.
For this purpose, a generalized Wannier equation approach which accounts for the mass anisotropy of L valley electrons is used~\cite{Springer:2016hz}.
This method is based on an expansion of the exciton wave function into spherical harmonics and introduces an energetic splitting of the 2p$_0$ and 2p$_{\pm 1}$ exciton states.
Effective electron masses and dielectric constants determine the extent of the splitting.
At a single L valley, the selection rules prohibit the 1s-2p$_0$ (1s-2p$_\pm 1$) transition except for purely parallel (perpendicular) polarization of the probe field with respect to the ellipsoidal axis of the conduction band.
However, the total THz response contains components from both polarizations due to the 4-fold degeneracy of the L valley in Ge.
As a result, the THz absorption is dominated by two distinct resonances.
Figure~\ref{fig:fig4} presents a comparison of the computed absorption (solid line) with the experimental spectrum 7\,ns after the optical excitation (shaded area).
The excellent agreement was obtained using effective L valley electron masses $m_{\perp}=0.05m_0$ and $m_{\parallel}=1.74m_0$, a hole mass of $m_\mathrm{h}=0.337m_0$, as well as a dielectric constant $\epsilon_\mathrm{r} = 16.2$.
In addition to the total spectrum, its decomposition into that part originating from parallel (dotted lines) and perpendicular (dashed lines) THz field polarization is shown.

In summary we performed optical pump and THz probe spectroscopy to study the dynamics of exciton formation in bulk Ge. 
The presence of excitons can be concluded from the occurrence of an absorption peak around  3.1\,meV (0.75\,THz) associated with the intra-excitonic 1s-2p transition. 
We find that exciton formation starts several hundred picoseconds after non-resonant excitation and attribute this delay to the fact that the carrier distribution needs to cool down first via acoustic phonons. Our experimental observations confirm the
theoretical prediction that exciton formation occurs faster for increased excitation densities.  
Furthermore, we show that the 1s-2p transition is composed of two distinct resonances which arise from the mass-anisotropy of L valley electrons. 

We acknowledge financial support from the Deutsche Forschungsgemeinschaft via the Collaborative Research Center 1083 (DFG:SFB1083).
\bibliography{references} 

\end{document}